\documentstyle[sprocl,epsf,bbox]{article}

\begin{document}

\title{Coulomb Final State Interactions for Gaussian Wave Packets}

\author{Urs Achim Wiedemann,}
\address{Physics Department, Columbia University, New York, NY 10027, USA}
\author{Daniel Ferenc and Ulrich Heinz}
\address{CERN/TH, CH-1211 Geneva 23, Switzerland, and \\
Institut f\"ur Theoretische Physik, Universit\"at Regensburg,
D-93040 Regensburg, Germany}

%%%%%%%%%%%%%%%%%%%%%%%%%%%%%%%%%%%%%%%%%%%%%%%%%%%%%%%%%%%%%%
% You may repeat \author \address as often as necessary      %
%%%%%%%%%%%%%%%%%%%%%%%%%%%%%%%%%%%%%%%%%%%%%%%%%%%%%%%%%%%%%%

\maketitle
\abstracts{ 
Two-particle like-sign and unlike-sign correlations including Coulomb 
final state interactions are calculated for Gaussian wave packets emitted
from a Gaussian source. We show that the width of the wave packets 
can be fully absorbed into the spatial and momentum space widths of an
effective emission function for plane wave states, and that Coulomb
final state interaction effects are sensitive only to the latter, but
not to the wave packet width itself. Results from analytical and
numerical calculations are compared with recently published work 
by other authors.
}

{\it 1. Introduction.} -- To analyze the geometry and dynamics of the
collision region, two-particle correlations $C(\bbox{q,K})$ of
like-sign and unlike-sign hadrons have been studied extensively in
relativistic heavy ion collisions at AGS~\cite{E877} and CERN
SPS~\cite{NA35,NA49} energies. They show $\bbox{q}$-dependent
structures at relative pair momenta $\vert\bbox{q}\vert < 100$
MeV. These originate mainly~\cite{GKW79} from (i) final state
interactions (which for pions at $\vert\bbox{q}\vert < 100$ MeV are
dominated completely by the Coulomb force) and (ii) the quantum
statistics of identical particles.  

Practical attempts to reconstruct space-time information from
two-pion correlation data in momentum space so far exploit mostly 
the quantum statistical Hanbury Brown -- Twiss (HBT) effect between
identical bosons~\cite{NA49,NA49R,WTH97}. This requires a prior
subtraction of final state Coulomb interaction effects from the
measured correlation functions, with proper account for the finite
size of the emission region~\cite{GKW79,P84,B91,F95,Baym96,AHR97}. Up to 
now this is done directly in the experimental analysis, either by
taking experimental unlike-sign pion correlations to correct the
like-sign ones~\cite{NA49}, or by a model calculation for the Coulomb
effect expected for a finite size emission region~\cite{E877}. 

This approach was recently questioned by Merlitz and
Pelte~\cite{MP97b,MP98a,MP98b}. From a numerical analysis based on
Gaussian wave packets emitted from a Gaussian source, they concluded
that~\cite{MP97b} ``the expected Coulomb distortion in the momentum  
correlation \dots becomes unobservable'' and that therefore
``experimental data, which are published after Coulomb correction, are
wrong for small momentum differences''. If correct, this conclusion
would invalidate a substantial part of the existing work on the
analysis of two-particle correlation data since it implies that either
(i) any attempt to base a space-time interpretation of identical
two-particle correlations on Coulomb corrected data is ill-founded or
that (ii) any attempt to describe the particle emitting source in
heavy ion collisions by a set of Gaussian wave packets is inconsistent. 

This dramatic perspective has led us to reconsider the calculation of
two-particle correlations for Gaussian wave packets. Following Merlitz and
Pelte we describe the particle emitting source in terms of a
distribution of wave packet centers and a characteristic wave packet
width $\sigma$ of the emitted particles. The emitted Gaussian wave
packets are propagated into the detector under the influence of mutual
Coulomb final state interactions. We derive analytical expressions
which show that the wave packet size $\sigma$ can always be absorbed
by a redefinition of the model parameters characterizing the source
size. For Gaussian source models, two-particle correlation measurements 
cannot 
%distinguish 
differentiate between ``source size'' and ``wave packet width''. 
In a limiting case we further prove analytically the equivalence of 
the Gaussian wave packet formalism and the usually adopted plane wave 
calculations {\it irrespective of the size of the wave packet width}. 
These analytical calculations show quite generally that the problem 
pointed out by Merlitz and Pelte does not exist. Their results disagree 
with our numerical calculations as well as with analytical formulae which 
we derive without approximations from the same starting point as the 
calculation presented in Ref.~\cite{MP97b}.  

%%%%%%%%%%%%%%%%%%%%%%%%%%%%%%%%%%%%%%%%%%%%%%%%%%%%%%%%%%%%%%%%%%%%%

{\it 2. Unlike-sign pion correlations.} -- Our starting point is a
set of Gaussian one-particle wave packets~\cite{MP97,W98,Weal97,CZ97}
  \begin{equation}
    f_i(\bbox{x},t_0) 
    = {1\over (\pi\sigma^2)^{3/4} }
    e^{-(\bbox{x}-\check{\bbox{r}}_i)^2/(2\sigma^2) 
    + i \check{\bbox{p}}_i\cdot\bbox{x} }\, ,
  \label{eq1}
  \end{equation}
which are centered at initial time $t = t_0$ at phase-space
points $(\check{\bbox{r}}_i,\check{\bbox{p}}_i)$. We expand the
time evolution of the corresponding two-particle state
$\Psi_{ij}(\bbox{x}_1,\bbox{x}_2,t_0)$ $= f_i(\bbox{x}_1,t_0)\, 
f_j(\bbox{x}_2,t_0) $ in terms of plane waves $\phi_{\bbox{p}_1,\bbox{p}_2}$,
  \begin{eqnarray}
    \Psi_{ij}(\bbox{x}_1,\bbox{x}_2,t) &=& 
    \int {d^3p_1\over (2\pi)^3}\, {d^3p_2\over (2\pi)^3}\,
    {\cal A}_{ij}(\bbox{p}_1,\bbox{p}_2,t)\, 
    \phi_{\bbox{p}_1,\bbox{p}_2}(\bbox{x}_1,\bbox{x}_2,t)\, ,
  \label{eq2} \\
    \phi_{\bbox{p}_1,\bbox{p}_2}(\bbox{x}_1,\bbox{x}_2,t) 
    &=& e^{-iEt}\, \phi_{2\bbox{K}}(\bbox{X})\, 
        \phi_{\bbox{q}/2}(\bbox{r})
    \equiv e^{-i(E_1+E_2)t}\, e^{2i\bbox{K}\cdot\bbox{X}}
      \, e^{{i\over 2}\bbox{q}\cdot\bbox{r}} ,
  \label{eq3} 
  \end{eqnarray}
which we write in terms of center of mass coordinates $\bbox{X} =
\textstyle{1\over 2}(\bbox{x}_1+\bbox{x}_2)$, $\bbox{K} =
\textstyle{1\over 2}(\bbox{p}_1+\bbox{p}_2)$, and relative coordinates
$\bbox{r} = (\bbox{x}_1-\bbox{x}_2)$, $\bbox{q} = (\bbox{p}_1 -
\bbox{p}_2)$. The probability ${\cal P}_{ij}$ for detecting at time
$t\to \infty$ the two particles prepared in the state $\Psi_{ij}$ with
momenta $\bbox{p}_1$ and $\bbox{p}_2$ is given by~\cite{AHR97} 
  \begin{eqnarray}
    {\cal P}_{ij}(\bbox{p}_1,\bbox{p}_2) &=&
    {\cal P}_{ij}(\bbox{q,K}) =
    \lim_{t\to \infty} \vert {\cal A}^*_{ij}(\bbox{p}_1,\bbox{p}_2,t)
    \vert^2 \, ,
  \label{eq4} \\
    \lim_{t\to \infty} {\cal A}_{ij}(\bbox{p}_1,\bbox{p}_2,t) 
    &=& \lim_{t\to \infty} 
    \langle e^{-i\hat{H}_0(t-t_0)} \phi_{\bbox{p}_1,\bbox{p}_2}(t_0)\,
    \vert\, e^{-i\hat{H}(t-t_0)} \Psi_{ij}(t_0) \rangle
    \nonumber \\
    &=& \langle \phi_{2\bbox{K}}\, \vert\, \Psi^{\rm pair}_{ij}\rangle\,
    \langle \Omega_+\, \phi_{\bbox{q}/2}\, \vert\,  
    \Psi^{\rm rel}_{ij}\rangle\, .
  \label{eq5}
  \end{eqnarray}
Here we separated the state $\Psi_{ij}(\bbox{x}_1,\bbox{x}_2,t_0)$
into relative and center of mass wave functions,
  \begin{eqnarray}
    \Psi^{\rm pair}_{ij}(\bbox{X}) &=& {1\over (\pi\sigma^2)^{3/4}}
    e^{-(\bbox{X}-\check{\bbox{X}}_{ij})^2/\sigma^2 
    + 2i\check{\bbox{K}}_{ij}\cdot\bbox{X} }\, ,
  \label{eq6}\\
    \Psi^{\rm rel}_{ij}(\bbox{r}) &=& {1\over (\pi\sigma^2)^{3/4}}
    e^{-(\bbox{r}-\check{\bbox{r}}_{ij})^2/(4\sigma^2) 
    + {i\over 2}\check{\bbox{q}}_{ij}\cdot\bbox{r} }\, ,
  \label{eq7}
  \end{eqnarray}
where $\check{\bbox{X}}_{ij}, \check{\bbox{K}}_{ij},
\check{\bbox{r}}_{ij}, \check{\bbox{q}}_{ij}$ are the corresponding
center-of-mass and
relative coordinates constructed from the wave packet centers.
In (\ref{eq5}) we have also introduced the M\o ller scattering
operator $\Omega_+ = \lim_{t\to\infty}$ $e^{i\hat{H}(t-t_0)}$  
${\rm e}^{-i\hat{H}_0(t-t_0)}$ for the final state interaction
hamiltonian $\hat{H} = \hat{H}_0 + V(\bbox{r})$, $\hat{H}_0 = - {1\over
  4 m}\Delta_ {\bbox{X}} - {1\over m} \Delta_{\bbox{r}}$. This M\o ller 
operator maps the plane wave $\phi_{\bbox{q}/2}$ onto the solution
of the Lippmann-Schwinger equation for the corresponding stationary
scattering problem. For two-particle Coulomb interactions this is
the Coulomb scattering wave
  \begin{eqnarray}
    \left(\Omega_+\, \phi_{\bbox{q}/2}\right)(\bbox{r}) &=&
    \Phi_{\bbox{q}/2}^{\rm coul}(\bbox{r}) = \Gamma(1-i\eta)\, 
    e^{{1\over 2}\pi\eta}\,
    e^{{i\over 2}{\bbox{q}\cdot\bbox{r}}}\, 
    F(i\eta; 1; iz_-)\, ,
  \label{eq8}\\
    z_\pm &=& {\textstyle{1\over 2}}(q r \pm \bbox{q}\cdot\bbox{r})\, ,
    \qquad
    \eta =  {m e^2\over 4\pi q}\, ,
  \label{eq9}
  \end{eqnarray}
where $e^2/ 4\pi = \alpha = 1/137$, $r = |\bbox{r}|$, $q = |\bbox{q}|$,
$F(i\eta; 1; iz_-)$ is the confluent hypergeometric function, and
$\eta$ is the Sommerfeld parameter. Eq.~(\ref{eq8}) applies for pairs
with opposite charges; for like-sign pairs one replaces $\eta\mapsto
-\eta$. With the help of Eq.~(\ref{eq8}), the calculation of the
amplitudes (\ref{eq5}) is reduced to a six-dimensional integral.   

%%%%%%%%%%%%%%%%%%%%%%%%%%%%%%%%%%%%%%%%%%%%%%%%%%%%%%%%%%%%%%%%%
 
{\it 2.1. Gaussian source model.} -- We consider a toy model of
simultaneous particle emission at time $t=t_0$ for which initially the
wave packet centers are distributed with Gaussians of
widths $R$ and $\Delta$ in coordinate and momentum space,
respectively. It can be specified by the following normalized
distributions of relative distances and pair coordinates: 
  \begin{eqnarray}
    S_{\rm rel}(\check{\bbox{r}},\check{\bbox{q}}) =
    {1\over (4\pi R \Delta)^3}
    e^{-{\check{\bbox{r}}^2\over 4R^2} 
       -{\check{\bbox{q}}^2\over 4\Delta^2}} ,
    \quad
    S_{\rm pair}(\check{\bbox{X}},\check{\bbox{K}}) =
    {1\over (\pi R \Delta)^3}
    e^{-{\check{\bbox{X}}^2\over R^2}
       -{\check{\bbox{K}}^2\over \Delta^2}} .
  \label{eq10}
  \end{eqnarray}
With the choice $R^2 = R_s^2/2$, $\Delta^2 = m\, T$, and $\sigma^2 =
2\, \sigma_0^2$, this model coincides with the one considered by
Merlitz and Pelte~\cite{MP97b}. We calculate the unlike-sign 
two-particle correlator via the two-particle spectrum ${\cal
  P}_{ij}(\bbox{p}_1, \bbox{p}_2)$ of Eq.~(\ref{eq4}), averaged over
the distributions (\ref{eq10}) and normalized to the
corresponding spectrum for pairs of non-interacting particles:
  \begin{eqnarray}
    && C^{+-}(\bbox{q,K}) = 
    {I^{\rm int}(\bbox{q,K})\over I^{\rm nonint}(\bbox{q,K})}\, ,
  \label{eq11} \\
    && I(\bbox{q,K}) = \int d^3\check{r}_{ij} d^3\check{q}_{ij} 
    d^3\check{X}_{ij} d^3\check{K}_{ij} 
    S_{\rm rel}(\check{\bbox{r}}_{ij},\check{\bbox{p}}_{ij}) 
    S_{\rm pair}(\check{\bbox{X}}_{ij},\check{\bbox{K}}_{ij})
    {\cal P}_{ij}(\bbox{q,K}) .
  \nonumber\\
  \label{eq12}
  \end{eqnarray}
For the non-interacting case Eq.~(\ref{eq12}) must be evaluated with
${\cal P}_{ij}^{\rm nonint}$ which is obtained by replacing in the
amplitude (\ref{eq5}) the relative Coulomb wave $\Omega_+\,
\phi_{\bbox{q}/2}$ by the plane wave $\phi_{\bbox{q}/2}$. 
Since the center of mass coordinate is not affected by the
two-particle final state interaction, the integrations over 
$S_{\rm pair}(\check{\bbox{X}}_{ij},\check{\bbox{K}}_{ij})$ drop out
in the ratio (\ref{eq11}), and the correlator $C^{+-}(\bbox{q,K})$
becomes independent of the pair momentum $\bbox{K}$. One finds
  \begin{eqnarray}
    &&C^{+-}(\bbox{q}) = G(-\eta) 
    \left( {\bar{\Delta}\over 4\pi\bar{R}}\right)^3\,
    e^{{\bbox{q}^2\over 4\bar{\Delta}^2}}\,
    \int d^3r\, e^{-{\bbox{r}^2\over 8\bar{R}^2} 
      + {i\over 2}\bbox{r}\cdot\bbox{q}}\,
      F(i\eta;1;iz_-)
  \nonumber \\
    && \qquad
    \times \int d^3r'\, e^{-{\bbox{r}'\over 8\bar{R}^2}
      - {i\over 2}\bbox{r}'\cdot\bbox{q}}\,
      [F(i\eta;1;iz'_-)]^*\,
      e^{-\left({\bar{\Delta}^2\over 4} - {1\over 16\bar{R}^2}
          \right)(\bbox{r}-\bbox{r}')^2} \, ,
  \label{eq13}
  \end{eqnarray}
where $G(\eta) = \vert \Gamma(1+i\eta) e^{-{1\over 2}\pi\eta}\vert^2 =
2\pi\eta/(e^{2\pi\eta}-1)$ is the Gamow factor. It is important that
this correlator depends only on the parameter combinations
  \begin{equation}
    \bar{R}^2 = R^2 + \textstyle{\sigma^2\over 2}\, , \qquad
    \bar{\Delta}^2 = \Delta^2 + \textstyle{1\over 2 \sigma^2}\, .
  \label{eq14}
  \end{equation}
This shows that for Gaussian models of particle emission, the wave packet
width $\sigma$ can be absorbed in a redefinition of the model parameters.
There is no measurement which allows to determine $\sigma$ independent
of $R$ and $\Delta$. Of course, the specific $\sigma$-dependence in
(\ref{eq14}) still constrains the values which $\bar{R}$ and
$\bar{\Delta}$ can take. In particular, $\bar R, \bar \Delta$ always
satisfy the uncertainty relation $\bar R \bar \Delta \geq \hbar/2$,
rendering the exponent of the last term in (\ref{eq13}) always
negative.  

 In the absence of final state interactions, the dependence of the 
correlator on the parameter combinations (\ref{eq14}) has been 
noted repeatedly~\cite{MP97,Weal97,Hcatania}. The momentum spectra
and correlations are entirely determined by the ``effective'' emission
function~\cite{Weal97,Hcatania}
  \begin{equation}
     S_{\rm eff}(\bbox{r},\bbox{p}) = \int d^3\check{r}\, d^3\check{p}\, 
     S(\bbox{\check{r},\check{p}})\, 
     S_{\rm w.p.}(\bbox{r-\check{r},p-\check{p}})\, ,
  \label{eq15}
  \end{equation} 
which is a folding integral between the distribution of wavepacket
centers $S(\check{\bf r},\check{\bf p})$ and the Wigner density
$S_{\rm w.p.}$ of a single particle wave packet. For Gaussian 
source models, $S_{\rm eff}$ depends on $\bar{R}$ and $\bar{\Delta}$
only~\cite{Hcatania}. The same holds true in the presence of final
state interactions where the correlator can be written for
arbitrary model distributions $S(\check{\bf r},\check{\bf p})$
as a quite involved expression depending on $S_{\rm eff}$ only
(see Eq. (60) in Ref.~\cite{AHR97}). Our result (\ref{eq13})
is an explicit representation of this general relation, obtained
for the Gaussian source models (\ref{eq10}). It allows further
analytical and numerical studies:

%%%%%%%%%%%%%%%%%%%%%%%%%%%%%%%%%%%%%%%%%%%%%%%%%%%%%%%%%%%%%%%%%

{\it 2.2. Limiting cases.} -- In two interesting limits the
correlator $C^{+-}(\bbox{q})$ can be further simplified analytically. 
Using $\lim_{\bar{\Delta}\to\infty}$ $({\bar{\Delta}^2/ 4\pi})^{3/2}$
$\exp\left[-{\bar{\Delta}^2\over 4}({\bf r}-{\bf r'})^2\right]$
$=\delta^{(3)}({\bf r}-{\bf r'})$ we find
  \begin{equation}
    \lim_{\bar{\Delta}\to\infty} C^{+-}(\bbox{q})
    = {1\over (4\pi\bar{R}^2)^{3/2}} \int d^3r\,
    e^{-{\bbox{r}^2\over 4\bar{R}^2}}\,
       \vert\Phi_{\bbox{q}/2}^{\rm coul}(\bbox{r})\vert^2 \, .
  \label{eq16}
  \end{equation}
This expression, first written down by Koonin~\cite{K77}, is the 
usually adopted starting point for plane wave 
calculations~\cite{E877,NA35,NA49,P84,B91,Baym96,SLPEA97,LL82,K77}; 
it was shown by Baym and Braun-Munzinger~\cite{Baym96} to be well
approximated by a semi-classical approach. The limit
$\bar{\Delta} \to \infty$ can be taken for arbitrary values of the
wave packet width $\sigma$ and is equivalent to the limit
$\Delta\to\infty$ which describes an emission function $S$ without
momentum dependence. In this limit, the correlator $C^{+-}(\bbox{q})$
for simultaneously emitted Gaussian wave packets coincides exactly
with the starting point of conventional plane wave
calculations~\cite{Baym96,AHR97}, {\it irrespective of the size
  $\sigma$ of the wave packet}. Rescaling $\sigma$ then simply amounts
to a change of the effective spatial size $\bar{R}$ of the source.

What happens if the source size becomes large? Changing in
Eq.~(\ref{eq13}) the integration variables $\bbox{r} \to
\sqrt{\bar{R}}\, \bbox{r}$, $\bbox{r}' \to \sqrt{\bar{R}}\, \bbox{r}'$,
and replacing the Coulomb wave function by its leading contribution,
$\Phi_{\bbox{q}/2}^{\rm coul}\left(\sqrt{\bar{R}}\bbox{r}\right)$
$\to \exp \left( {i\over 2}\sqrt{\bar R} \bbox{r}\cdot\bbox{q} \right. +$
$\left. i\eta\ln \left({\sqrt{R}\over 2}(qr-\bbox{q}\cdot\bbox{r})\right)
\right)$ $+ O\left({1\over \sqrt{R}}\right)$, we find
  \begin{eqnarray}
    \lim_{\bar{R}\to \infty}\, C^{+-}(\bbox{q}) &=& 1\, .
  \label{eq17}
  \end{eqnarray}
This is expected: as the source becomes larger, the average spatial
separation between particles increases and their Coulomb
attraction decreases, leading to a flat correlator in the limit of
infinite source size.

\begin{figure}[ht]\epsfxsize=7.5cm 
 \centerline{\epsfbox{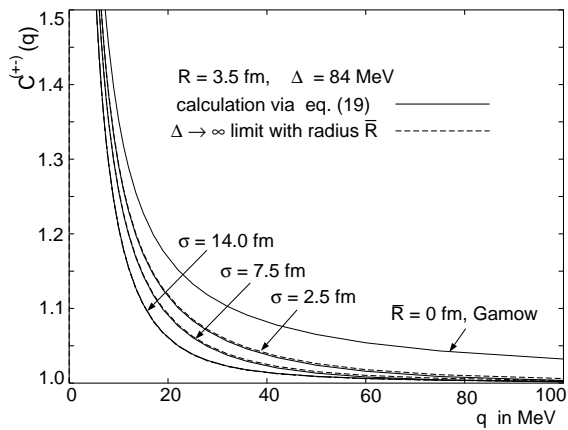}}
 \caption{Two-particle correlator of unlike-sign pion pairs for the
   Gaussian model (\ref{eq10}). The correlation depends
   only on the model parameter combinations $\bar{R}^2 = R^2 +
   \sigma^2/2$ and $\bar{\Delta}^2 = \Delta^2 + 1/(2\sigma^2)$. The
   results for $\Delta = 84$ MeV agrees well with Koonin's expression
   (\ref{eq16}) obtained in the limit $\Delta \to \infty$. 
}\label{fig1}
 \end{figure}

{\it 2.3. Numerical results.} -- One may wonder whether a large but
realistic effective source size $\bar{R}$ can come sufficiently close
to the limiting case $\bar{R} \to \infty$ of (\ref{eq17}) to support
the claim of Merlitz and Pelte~\cite{MP97b,MP98a} that the Coulomb
repulsion becomes effectively unobservable. To study this question we
have calculated the correlator (\ref{eq13}) numerically, after doing 
the azimuthal integrations ($q=\vert\bbox{q}\vert$):
  \begin{eqnarray}
    C^{+-}(q) &=& 4\pi^2\, G(-\eta)
    \left( \bar{\Delta}\over 4\pi\bar{R}\right)^3
    e^{{q^2\over 4\bar{\Delta}^2}}
    \nonumber \\
    &\times& \int_0^\infty r^2\, dr\, {\rm e}^{-{r^2\over 8\bar{R}^2}}
    \int_{-1}^1 dx\, {\rm e}^{{i\over 2}qrx} 
    F(i\eta;1;\textstyle{i\over 2}qr(1-x))
    \nonumber \\
    &\times& \int_0^\infty r'^2\, dr'\, {\rm e}^{-{r'^2\over 8\bar{R}^2}}
    \int_{-1}^1 dy\, e^{-{i\over 2}qr'y} 
    [F(i\eta;1;\textstyle{i\over 2}qr'(1-y))]^*
    \nonumber \\
    &\times& I_0\left(2\, B^2\, rr'\sqrt{1-x^2}\sqrt{1-y^2}\right)\, 
    e^{-B^2 (r^2 - 2rr'xy + r'^2)}\, .
    \label{eq18}
  \end{eqnarray}
Here $I_0$ is the modified Bessel function and $B^2 = \bar{\Delta}^2/4
+ 1/(16\bar{R}^2)$. The numerical results for $C^{+-}(\bbox{q})$ 
are shown in Fig.~\ref{fig1} for the model parameters $R = 3.5$ fm, 
$\Delta = 84$ MeV and different values of the wave packet width $\sigma$. 
For $\sigma = 2.5$ fm, these values correspond to the model parameters 
chosen in Ref.~\cite{MP97b}.

%%%%%%%%%%%%%%%%%%%%%%%%%%%%%%%%%%%%%%%%%%%%%%%%%%%%%%%%%%%%%%%%%%%%
%%%%%%%%%%%%%%%%%%%%%%%%%%%%%%%%%%%%%%%%%%%%%%%%%%%%%%%%%%%%%%%%%%%%%%%

One sees that, with increasing source size $\bar{R}^2 = R^2
+\sigma^2/2$, the resulting Coulomb correlation indeed becomes smaller
than the Gamow factor, but it remains clearly observable even for a
very large wave packet size $\sigma = 14$ fm. We also compare in
Fig.~\ref{fig1} the full correlator $C^{+-}(q)$ of
Eqs.~(\ref{eq13}/\ref{eq18}) to the limit (\ref{eq16}), calculated for
the same value of $\bar{R}$. At least for the model parameters studied
here, both expressions agree almost exactly. There seems to be no
possibility to distinguish on the basis of final state correlations
between the Gaussian wave packet formalism and the plane wave
calculation based on the approximate expression (\ref{eq16}) which was
used in other studies~\cite{E877,NA49,Baym96}. As long as the Gaussian
wave packet width $\sigma$ is included consistently in the definition
of the source size $\bar{R} = \sqrt{R^2 + \sigma^2/2}$, both
formalisms lead qualitatively and quantitatively to the same
result. We conclude that the differing results derived in
Ref.~\cite{MP97b} with the help of a numerical simulation of the time
evolution of wave packets are incorrect.

%%%%%%%%%%%%%%%%%%%%%%%%%%%%%%%%%%%%%%%%%%%%%%%%%%%%%%%%%%%%%%%%%%%%%
{\it 3. Like-sign pion correlations.} -- For pairs of pions of
identical charge the Coulomb final state effects are
superimposed on the quantum statistical effects resulting
from the symmetrization of the two-particle wave function. 
Paralleling the calculation of section 2. with
Bose-Einstein symmetrized Gaussian wavepackets $\Psi_{ij}$
and plane waves $\phi_{\bbox{p}_1,\bbox{p}_2}$, one arrives
at the symmetrized asymptotic two-particle amplitude
  \begin{eqnarray}
%    \!\!\!\!
    \lim_{t\to\infty} {\cal A}^{\rm BE}_{ij}(\bbox{p}_1,\bbox{p}_2,t) 
    = \langle \phi_{\bf K}\, \vert\, \Psi^{\rm pair}_{ij}\rangle\,
    \left[ \langle \Omega_+\, \phi_{\bbox{q}/2}\, \vert\,  
    \Psi^{\rm rel}_{ij}\rangle + \langle \Omega_+\,
    \phi_{-\bbox{q}/2}\, \vert\,  
    \Psi^{\rm rel}_{ij}\rangle \right] ,
  \label{eq19}
  \end{eqnarray}
from which the two-particle momentum-space probability ${\cal
  P}_{ij}^{\rm BE} (\bbox{p}_1,\bbox{p}_2)$ is again calculated
according to Eq.~(\ref{eq4}). Averaging ${\cal P}_{ij}^{\rm BE}$
according to (\ref{eq12}) over the model distribution gives the
numerator of the two-particle correlator which we call $I^{\rm
  BE,int}(\bbox{q,K})$. We normalize it by the method of ``mixed
pairs'': an uncorrelated (mixed) pair is described by an unsymmetrized
product state without Coulomb interaction and leads, after averaging
over the model distribution, to $I^{\rm nonint}(\bbox{q,K})$ (see
Eqs.~(\ref{eq11}/\ref{eq12})). Taking both distinguishable states
$f_i(\bbox{x}_1,t_0)\, f_j(\bbox{x}_2,t_0)$ and $f_i(\bbox{x}_2,t_0)\,
f_j(\bbox{x}_1,t_0)$ into account we have
  \begin{eqnarray}
    C^{++}(\bbox{q,K})={I^{\rm BE, int}(\bbox{q,K}) \over
    I^{\rm nonint}(\bbox{q,K}){+}I^{\rm nonint}({-}\bbox{q,K})}
    = C^{\rm dir}(\bbox{q}) + C^{\rm ex}(\bbox{q})\, .
  \label{eq20}
  \end{eqnarray}
For the Gaussian model (\ref{eq10}), the center of mass coordinate
is affected neither by two-particle final state interactions nor by
two-particle Bose-Einstein symmetrization. The correlator (\ref{eq20})
hence does not depend on the pair momentum $\bbox{K}$. It
splits into two contributions. The ``direct term'' 
can be obtained from $C^{+-}(\bbox{q})$ in Eq.~(\ref{eq13}) by
changing the sign of the Sommerfeld parameter, $-\eta \to \eta$. 
The ``exchange term'' is given by 
  \begin{eqnarray}
    &&C^{\rm ex}(\bbox{q}) = {\rm Re}\,\bigg\lbrace G(\eta) 
    \left( {\bar{\Delta}\over 4\pi\bar{R}}\right)^3\,
    e^{{\bbox{q}^2\over 4\bar\Delta^2}}\,
    \int d^3r\, e^{-{\bbox{r}^2\over 8\bar{R}^2}
      + {i\over 2}\bbox{r}\cdot\bbox{q}}\, F(-i\eta;1;iz_-)
    \nonumber \\
    && \qquad
    \times \int d^3r'\, e^{-{\bbox{r}'^2\over 8\bar{R}^2}
      + {i\over 2}\bbox{r}'\cdot\bbox{q}}\, [F(-i\eta;1;iz'_+)]^*\,
      e^{ -\left( {\bar{\Delta}^2\over 4} - {1\over 16\bar{R}^2}
           \right) (\bbox{r}-\bbox{r}')^2}\bigg\rbrace\, .
  \label{eq21}
  \end{eqnarray}
This integral can be simplified to a 4-dimensional expression similar
to (\ref{eq18}). The limits $\bar\Delta\to\infty$ and $\bar R\to
\infty$ of the first term $C^{\rm dir}(\bbox{q})$ are obtained from
(\ref{eq16}/\ref{eq17}) by replacing $-\eta \to \eta$. The corresponding
limits for the exchange term are given by  
  \begin{eqnarray}
    \lim_{\bar{\Delta}\to\infty} C^{\rm ex}(\bbox{q})
    &=& {1\over (4\pi\bar{R}^2)^{3/2}} \int d^3r\,
    e^{-{\bbox{r}^2\over 4\bar{R}^2}}\, \cos(\bbox{r}\cdot\bbox{q})\,
    \vert \Phi^{\rm coul}_{\bbox{q}/2}(\bbox{r})\vert^2 \, ,
  \label{eq22}\\
    \lim_{\bar{R}\to \infty}\, C^{\rm ex}(\bbox{q}) &=& 0\, .
  \label{eq23}
  \end{eqnarray}
As in the case of unlike-sign correlations, the correlator
$C^{++}(\bbox{q})$ depends only on the parameter combinations
$\bar{R}$ and $\bar{\Delta}$, but not explicitly on the  
wave packet width $\sigma$. In the limit $\bar{\Delta} \to \infty$,
the Gaussian wave packet formalism again coincides with the
Koonin formula~\cite{K77} (now with the additional symmetrization
factor $1 + \cos(\bbox{q}\cdot\bbox{r})$ under the integral) which is
the starting point of most conventional plane wave calculations.  

%%%%%%%%%%%%%%%%%%%%%%%%%%%%%%%%%%%%%%%%%%%%%%%%%%%%%%%%%%%%%%%%%%%%
 \begin{figure}[ht]\epsfxsize=7.5cm 
 \centerline{\epsfbox{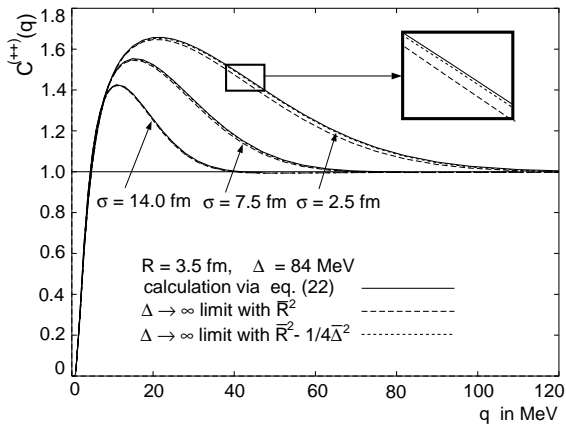}}
 \caption{Two-particle correlator of like-sign pion pairs for the
   Gaussian model (\ref{eq10}). The correlation depends
   only on the model parameter combinations $\bar{R}^2 = R^2 +
   \sigma^2/2$ and $\bar{\Delta}^2 = \Delta^2 + 1/(2\sigma^2)$. The
   full correlator is well approximated by the standard Koonin
   expression for a static momentum-independent source of radius
   squared $\bar{R}^2 - 1/(4\bar{\Delta}^2)$.
%   $\sqrt{\bar{R}^2 - 1/(4\bar{\Delta}^2)}$.
  } 
\label{fig2}
 \end{figure}
%%%%%%%%%%%%%%%%%%%%%%%%%%%%%%%%%%%%%%%%%%%%%%%%%%%%%%%%%%%%%%%%%%%%%%%

In Fig.~\ref{fig2} we compare numerically the full correlator
$C^{++}(q)$ from Eq.~(\ref{eq20}), for $\Delta = 84$ MeV,
with the limit $\Delta \to \infty$, both calculated for the same value
of $\bar{R}^2 = R^2 + \sigma^2/2$. (Due to spherical symmetry of the
source the correlator depends only on $q=\vert\bbox{q}\vert$.) For
small values of $\sigma$ one observes a small, but significant
difference. The reason is that even in the absence of Coulomb final
state interactions, the HBT radius parameter (which gives the
$q$-width of the correlator) is not exactly given by the source size
$\bar{R}^2$ but rather by~\cite{Weal97}   
  \begin{equation}
    R_{\rm HBT}^2 = \bar{R}^2 - {1\over 4\bar{\Delta}^2}\, .
  \label{eq24}
  \end{equation}
For large values of $R$ or $\sigma$, the term $\bar{R}^2$ dominates
this expression, and the difference between $R_{\rm HBT}^2$ and
$\bar{R}^2$ disappears (see Fig.~\ref{fig2}). In fact, when
computing the limit $\Delta \to \infty$ of $C^{++}(q)$ using 
$R_{\rm HBT}^2$ instead of $\bar{R}^2$, the agreement with the full
correlator $C^{++}(q)$ becomes almost exact even for small values of
$\sigma$ (see inset in Fig.~\ref{fig2}). We note that even for the
smallest value studied here, $\sigma = 2.5$ fm, the term
$1/(4\bar{\Delta}^2)$ contributes only $\approx 5$\% to $R_{\rm
  HBT}^2$. This small difference is clearly visible in the
exchange term (\ref{eq21}) of $C^{++}(q)$, whereas its influence on
the direct term $C^{\rm dir}$ (and hence on the correlator $C^{+-}$)
is found numerically to be an order of magnitude smaller. This 
illustrates that the two-particle correlator $C^{++}$ of identical 
pions is more sensitive than $C^{+-}$ to a small change in the 
Gaussian width of the phase space density.

The modification (\ref{eq24}) of the radius parameter to be used as 
input in the plane wave calculation can also be obtained from the
Koonin expression~\cite{K77} or, most explicitly, from Eq.~(65) of
Ref.~\cite{AHR97}. The consistency of Koonin's expression with the
full correlator in the present model calculation is a non-trivial 
check of the so-called smoothness approximation used in
Ref.~\cite{AHR97} to derive Koonin's expression from a general
treatment of two-body final state interactions. 

To sum up: as long as the Gaussian
wave packet width $\sigma$ is included consistently in the definition
of the source size, both the plane wave 
calculations~\cite{E877,NA35,NA49,P84,B91,Baym96,SLPEA97,LL82,K77} 
and the Gaussian wave packet formalism~\cite{MP97,W98,Weal97,CZ97} 
lead to qualitatively and quantitatively
equivalent results. While the present study proved this only for
Gaussian source models, we expect it to be true quite generally
since we know that~\cite{AHR97} the relation (\ref{eq15}) between 
the effective emission function and the Wigner density of single
particle wave packets holds for arbitrary model distributions 
and that~\cite{He96} two-particle momentum correlations
are mostly sensitive to the Gaussian characteristics of the
source in space-time.

%%%%%%%%%%%%%%%%%%%%%%%%%%%%%%%%%%%%%%%%%%%%%%%%%%%%%%%%%%%%%%%%%%%%%%% 
{\it Acknowledgments:}
This work was supported by the U.S. Department of Energy under Contract
No. DE-FG02-93ER40764, by DFG, BMBF and GSI. We thank P.~Braun-Munzinger, 
H.~Feldmeier, M.~Gyulassy, J.~H\"ufner, and J.~Stachel for stimulating 
discussions, and D.~Pelte for an open debate of 
Refs.~\cite{MP97b,MP98a,MP98b}.

\end{document}